\documentclass{svmult}

\usepackage[sort&compress,numbers]{natbib}

\usepackage{makeidx}         
\usepackage{graphicx}        
\usepackage{multicol}        
\usepackage[bottom]{footmisc}

\makeindex             

\begin{document}

\title*{Dynamics in online social networks}
\author{Przemyslaw A. Grabowicz, Jos\'e J. Ramasco, V\'{\i}ctor M. Egu\'{\i}luz}
\authorrunning{Grabowicz et al}
\institute{Instituto de F\'{\i}sica Interdisciplinar y Sistemas Complejos IFISC (CSIC-UIB), E07122 Palma de Mallorca, Spain\\
Email: \texttt{pms@ifisc.uib-csic.es}
}
%
%
\maketitle

\section{Introduction}

An increasing number of today's social interactions occurs using online social media as communication channels. Some online social networks have become extremely popular in the last decade. They differ among themselves in the character of the service they provide to online users. For instance, Facebook can be seen mainly as a platform for keeping in touch with close friends and relatives, Twitter is used to propagate and receive news, LinkedIn facilitates the maintenance of professional contacts, Flickr gathers amateurs and professionals of photography, etc. Albeit different, all these online platforms share an ingredient that pervades all their applications. There exists an underlying social network that allows their users to keep in touch with each other and helps to engage them in common activities or interactions leading to a better fulfillment of the service's purposes. This is the reason why these platforms share a good number of functionalities, {\it e.g.}, personal communication channels, broadcasted status updates, easy one-step information sharing, news feeds exposing broadcasted content, etc. As a result, online social networks are an interesting field to study an online social behavior that seems to be generic among the different online services.
Since at the bottom of these services lays a network of declared relations and the basic interactions in these platforms tend to be pairwise, a natural methodology for studying these systems is provided by network science. In this chapter we describe some of the results of research studies on the structure, dynamics and social activity in online social networks. We present them in the interdisciplinary context of network science, sociological studies and computer science.

\section{Structure of social networks}

Social networks in general show a very rich internal structure \cite{Newman2003Why} that in some aspects falls quite far from random graphs or even from artificial networks created by virtue of a preferential attachment mechanism. In this section we briefly review the most important features broadly found in social networks.

\subsection{Degree distribution}
The most fundamental characteristic of a network is the distribution of degrees: a function that measures how many friends have the members of the network and what is the variability of this number among all the users. The degree distribution in social networks are usually broad. These distributions have been typically modeled as functions having a heavy tails such as a power-law or  log-normals combined with an exponential cut-off at large values of the number of friends \cite{Ahn2007Analysis, Boccaletti2006Complex, Leskovec2008Planetary-scale, Mislove2007Measurement, Newman2003structure, Ugander2011Anatomy}. This means that there is a large variability in the number of connections of the nodes, with many nodes having small or moderate number of friends and a small number of them maintaining large number of friends. Almost all users of online social networks are connected in a largest connected component \cite{Leskovec2008Planetary-scale, Ugander2011Anatomy}. Some of the studies also point out that online social networks contain a densely connected core or cores \cite{Leskovec2008Planetary-scale,Mislove2007Measurement} consisting in groups of high-degree nodes that hold the network together. The existence of such cores provides paths for the connection between distinct parts of the network. A well-known aspect of the social networks is that the average shortest-path distance is low \cite{Ahn2007Analysis, Leskovec2008Planetary-scale, Mislove2007Measurement, Ugander2011Anatomy}. This characteristic is popularly known as six degrees of separation or small-world effect \cite{Travers1969experimental}. The importance of the shortcuts for reducing the network path-length has been highlighted in Ref.~\cite{Watts1998Collective}.

\subsection{Triangles and community structure}
Possibly, the most important feature distinguishing social networks from other types of networks is their high level of clustering or transitivity \cite{Ahn2007Analysis, Leskovec2008Planetary-scale, Mislove2007Measurement, Newman2003structure, Newman2003Why, Watts1998Collective,  Ugander2011Anatomy}. The clustering coefficient measures the probability that two nodes sharing a common neighbor (a node to which they are both connected) are connected. This property is quantified with a global clustering coefficient $C$ \cite{Newman2003structure} which is defined as
\begin{equation}
C=\frac{\textnormal{number of closed connected triples}}{\textnormal{number of connected triples}},
\end{equation}
where a connected triple of nodes is a sequence of $3$ nodes which have at least $2$ connections between them, and a closed triple is a triangle. One can also define a local clustering coefficient $c_i$ as:
\begin{equation}
c_i=\frac{\textnormal{number of closed triples centered on node i}}{\textnormal{number of triples centered on node i}}.
\end{equation}
In this case a global value of clustering coefficient may be obtained averaging the local $c_i$ over all the nodes of the network $\langle c \rangle$. One should note that $\langle c  \rangle$  is in general different from the coefficient $C$, and that the latter has a much worse scaling behavior.
At the structural level, a high clustering coefficient indicates the presence of many triangles in the network. At the social level, this means that friends of an individual tend to be connected between themselves too. This is a well-known phenomena in sociology which is important for the formation of strong social ties \cite{Krackhardt2007Heider, Granovetter1973strength} and affects to the emergence of positive and negative relations \cite{Leskovec2010Predicting}. At the network macroscopic level, a high density of  triangles can be related to the existence of community structure in social networks \cite{Burt2005Brokerage}. Furthermore the study \cite{Foster2011Clustering} suggests that in real networks with high value of clustering coefficient community structure emerges without any additional ingredients included.

Existence of communities in social networks is considered to have high relevance both by sociologist \cite{Burt2005Brokerage, Granovetter1973strength} and network scientists \cite{Fortunato2010Community, Grabowicz2012Social, Palla2007Quantifying}. We give further arguments on this in the third section of this chapter. In online social networks, groups can be identified in several ways. One of them is searching for communities in the graph defined as more densely connected parts of the network compared with their neighborhood. This approach is usually taken in network science and various community detection algorithms have been developed and continue to be under active development for detecting such clusters \cite{Fortunato2010Community}. In addition, some online social networks allow their users to create explicit groups and to claim its membership. Although it seems straightforward to make use of such user-declared groups, one should be careful when interpreting them since incentives for creation of such groups may vary \cite{Pissard2007Thematic}. Nevertheless it has been found that declared groups tend to have internally higher clustering coefficient \cite{Mislove2007Measurement} and therefore they may be correlated with the more-densely connected parts of the network found by community detection algorithms.

\subsection{Assortativity and homophily}
Another common feature of social networks is that connected users tend to be similar \cite{Mcpherson2001Birds}. This effect is popularly known as \textit{birds of a feather flock together} phenomena. It manifests itself in social networks through similarities in various properties of connected individuals. From pure network theory point of view the similarity may appear as a correlation of degrees between friends, which is called assortativity mixing, or as a rich-club effect \cite{Opsahl2008Prominence}. In such assortative networks nodes of high degrees tend to be connected to other nodes of high degrees, and vice versa, nodes of low degree tend to be connected to other low degree nodes. It has been found that offline social network are assortative  in contrast with networks of other types \cite{Newman2002Assortative, Newman2003Why}. However, this is not the only property in which friends are similar. This kind of phenomena is in general called homophily and is known to be present very broadly is social networks. People who are connected in online social networks tend to have similar age, live in close locations, and have similar interests \cite{Palla2007Quantifying, Ugander2011Anatomy, Leskovec2008Planetary-scale, Schifanella2010Folks}. It is also thought that people who belong to the same community, namely the same well-connected group of people, talk about similar topics, which can have an important impact on information and innovation diffusion in social networks \cite{Granovetter1973strength,Centola2011Experimental}.

\subsection{Differences between offline and online social networks}
As shown in the previous subsections many statistical properties of offline social networks are also found online. On the other hand creating links in a social online network is much less costly than developing offline social relations. These online connections can easily accumulate and pile up to large numbers~\cite{Avnit2009Million}. If the number of connections increases to the millions, the amount of effort that a user can invest into a relation that each link represents must fall to near zero. An early illustration of the relevance of the definition of social tie in characterization of social networks was shown in the study of email networks: while the distribution of the number of contacts in address books is power law \cite{Ebel2002Scale-free}, it is exponential when the contacts are restricted to reciprocated emails \cite{Guimera2003Self-similar}. Moreover disassortative mixing has been encountered in some online networks \cite{Ahn2007Analysis,Hu2009Disassortative} in contrast to the assortative mixing characteristic of offline social networks \cite{Newman2003structure}. As a matter of fact there exists an open discussion on the validity of online interactions as indicators of real social activity~\cite{Avnit2009Million,Cummings2002quality,Lazer2009Computational,Vespignani2009Predicting,Watts2007twenty-first}. In order to test the validity of online networks for social studies and to find its limitations further investigation is needed. In this chapter we present some of recent results of such studies.

\section{Growth in social networks}

\subsection{Preferential attachment}

Many features of complex systems are characterized by  \mbox{heavy-tailed} distributions~\cite{Newman2005Power, Saichev2009Theory}, {\it e.g.}, frequency of words~\cite{Zipf1949Human}, the wealth of nations~\cite{Pareto1964Cours}, degree distribution of complex networks~\cite{Barabasi1999Emergence}, etc. This property is typically perceived as a symptom of the rich-gets-richer principle, from which the so-called preferential growth models stem. The common concept of these models is that the elements of the system grow proportionally to their current size, what is referred to as preferential growth or preferential attachment rule. Typically, in these models, increments  of the defining variables of the system occur in each time step. Such increments can involve the addition of  new elements and/or to increase the sizes of the existing ones according to a preferential growth rule. Preferential models are usually the first approach to explain \mbox{heavy-tailed} distributions in many different systems~\cite{Simon1955class, Eisenberg2003Preferential, Yamasaki2006Preferential, Maruvka2011Birth-Death-Mutation}. In the case of networks, this kind of models got popularized a decade ago \cite{Barabasi1999Emergence, Barabasi1999Mean-field, Huberman1999Internet:, Dorogovtsev2000Structure, Bornholdt2001World}. The first of these models in the context of complex networks was introduced by Barab\'asi and Albert in Ref.~\cite{Barabasi1999Mean-field}. To describe it shortly: in each time step one node is introduced to the system with $m$ edges. These edges are connected to existing nodes in the system with probability proportional to the degree of the present nodes. As a result a network with \mbox{heavy-tailed} (usually \mbox{power-law}) distribution of node degrees emerges. The rules of Barabasi's model yield high simplicity, which is typically  a desirable feature, but that can be too rigid in some cases. In preferential-growth models the time unit is directly coupled to the number of new arriving elements, which can complicate the comparison of the dynamics of these models with real data. Some other drawbacks include the lack of heterogeneity and strong correlation between age of elements and their size~\cite{Adamic2000Power-Law}. Because of these issues the basic preferential growth model is typically used as a simple \mbox{toy-model} for generation of networks with \mbox{power-law} distribution of degrees. On the other hand, it is also successfully used as a component of models trying to simulate growth of real social networks \cite{Mislove2008Growth, Leskovec2008Microscopic}.

\subsection{Heterogeneity}

In many real systems, especially in social systems, individuals or elements are very diverse. This factor is related to the  \mbox{heavy-tailed} distributions that are so commonly found. In this direction, some models incorporating heterogeneity in the form of fitness, hidden variables or ranking shuffling have been proposed~\cite{Bianconi2001Competition, Caldarelli2002Scale-Free, Soderberg2002General, Boguna2003Class, Ratkiewicz2010Characterizing}. In general this family of models determines growth of elements with some kind of intrinsic property. Whereas in preferential attachment models the growth was proportional to current size of the elements, in fitness models it is usually proportional to the intrinsic fitness of each element. Typically the fitness is a random variable specific for each element drawn from a given variables. This introduces high heterogeneity between the elements. A number of empirical works shows how this intrinsic fitness is distributed and what is its role in complex system growth~\cite{Garlaschelli2004Fitness-Dependent, DeMasi2006Fitness, Kong2008Experience, Grabowicz2012Heterogeneity}. We discuss in detail one of the models of this family in the next section when commenting on the growth of groups in online social networks.

\subsection{Triadic closure / Triangle closing}

Due to the fact that clustering coefficient is remarkably high in some networks (mainly social networks), other growth models have been introduced in order to reproduce high number of triangles in the network. One of the first models accounting for this was \cite{Watts1998Collective} in which regular network with initially high clustering had its connections rewired to make it more realistic and control clustering coefficient, as well as average shortest path length. A more sophisticated model used to simulate growth of social networks has been proposed in Ref.~\cite{Leskovec2008Microscopic} and one of its main components is triangle closing. In this model new nodes appearing in the system connect to some node, usually using preferential attachment rule, and then start closing triangles with neighbors of this node. This simple triangle closing mechanism exhibit  much more realistic results in modeling online social networks \cite{Leskovec2008Microscopic}.

\begin{figure}\begin{center}
\includegraphics[width=12cm]{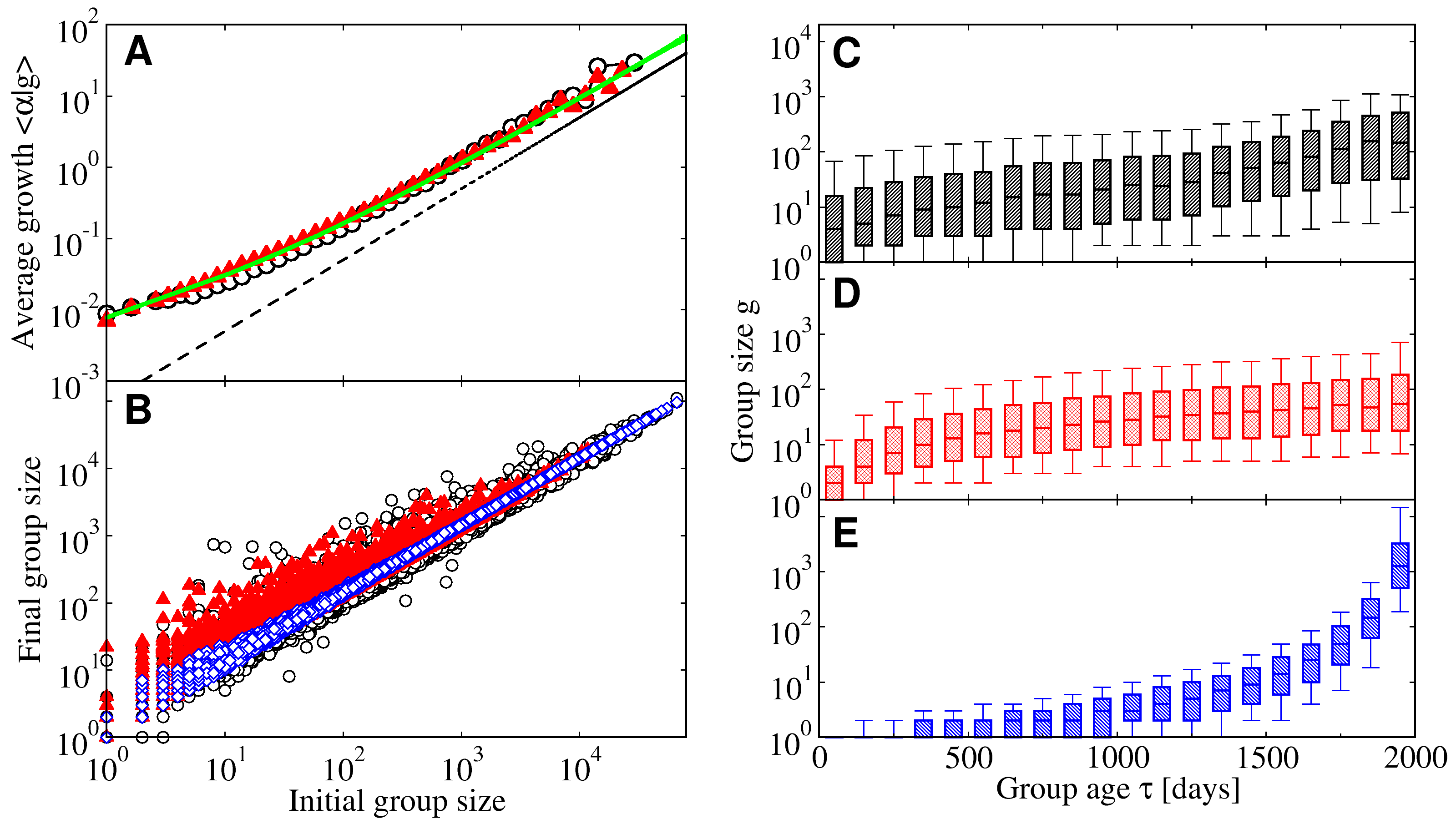}
\caption[ ]{Growth of groups in Flickr. (A) Average daily growth as a function of the initial size of the groups, estimated for the period of 6 weeks and averaged over 260,000 groups of a given initial size, for:  the real data from Flickr (circles), the heterogeneous linear growth model (triangles). The dashed line corresponds to the linear behavior $\langle \alpha | g\rangle \sim g$. (B) Initial and final group sizes over a period of 350 days for the real data (circles), the heterogeneous linear growth model (filled triangles) and Simon model (diamonds). Each point represents a single group, there are 9,503 points plotted for each set of points. (C-E) Box plots with whiskers at 9th/91st percentile of final size of groups as a function of their age at the time of the measurement for 260,000 groups for (C) the real data, (D) the heterogeneous linear growth model, (E) the Simon model. Adapted from Ref.~\cite{Grabowicz2012Heterogeneity}.} \label{fig_flickrscomparison}
\end{center}\end{figure}

\subsection{Dynamics of groups}

As we have emphasized in the previous section the existence of communities plays an important role in functioning of social networks. In this section we present studies of the growth of such groups. Several aspects have been identified as positively influencing groups' growth and their persistence. It has been suggested that high internal connectivity helps declared groups' growth \cite{Taraborelli2011Viable}. Other work argues that flexibility of big communities helps them stay alive longer, while small groups are more persistent if their composition stays unchanged \cite{Palla2007Quantifying}.

From the macroscopic perspective growth of groups can be described and modeled using a version of preferential attachment model or a model with hidden variables/intrinsic heterogeneity.
A comparison between these two approaches has been performed in Ref.~\cite{Grabowicz2012Heterogeneity} using real data from Flickr. The heterogeneous linear growth model suggested in this study assumes linear growth of groups with growth value (fitness) being drawn from heavy-tailed distribution (log-normal) and a number of new groups appearing in the systems growing linearly in time. As a comparison, a version of Simon model \cite{Simon1955class} has been used, which represents a model from preferential attachment family. As one can see in Figure~\ref{fig_flickrscomparison}A, the  average growth $\langle \alpha | g \rangle $ for groups of given size $g$ is proportional to the size of the groups for high $g$. This commonly is interpreted as the consequence of preferential attachment. However, as it is shown in Figure~\ref{fig_flickrscomparison}A, one obtains similar dependence using the heterogeneous linear growth model. This is the case because the average growth is an average over all groups of a given size, each of them growing linearly. Due to the heterogeneity and the linear growth, at a given time larger groups consist of old groups that grow slowly and younger groups that grow faster. Thus, the observation of preferential growth for groups of the same size does not reflect in this case an underlying rich-gets-richer principle, but it is a consequence of the competition of groups with different growth values and ages. Both the heterogeneous linear growth model and the Simon model produce \mbox{heavy-tailed} distribution of group sizes. However, the former model performs better in other respects. First, in the Simon model the final size of groups is heavily determined by their initial size measured one year before (Figure~\ref{fig_flickrscomparison}B), thus there is little heterogeneity among the groups, in contrast to the heterogeneous linear growth model which displays a degree of heterogeneity similar to the one of real groups. Second, for the Simon model the correlation of size and age is strong, while it is weak for real groups and the heterogeneous linear growth model (Figure~\ref{fig_flickrscomparison}C-E).
In the heterogeneous linear growth model the \mbox{heavy-tailed} distribution of final sizes of elements does not emerge from the growth process itself \mbox{(e.g., rich-gets-richer principle)}, but from the intrinsic heterogeneity of elements which take part in this growth process. This certainly does not answer the question why some groups grow faster than the others, as we do not understand yet what factors influence the fitness of the groups. However it points that it does not have to be due to the fact that one group is bigger than the other as in preferential attachment models. The simplicity of this approach suggests that the characterization of the heterogeneity may play an important role in understanding the origin of broad distributions and the time evolution of many real systems.

\section{Activity in the online social networks}
In general a social network is a broad term and it refers to a set of actors and a set of ties between them representing some kind of relation or interaction. In fact, however, there are many types of both relations and interactions, and usually they happen on top of each other. So far we mostly discussed social networks which represent a particular relation or interaction, e.g.: co-appearance in movies, boards of directors or co-authorship \cite{Newman2003Why,Newman2002Assortative,Newman2003structure}, network of online friendship \cite{Mislove2007Measurement,Ahn2007Analysis, Leskovec2008Microscopic, Ugander2011Anatomy}, network of communication \cite{Onnela2007Structure, Palla2007Quantifying, Leskovec2008Planetary-scale}. In online social networks, we can relate user activity with their declared relations with other users. In other words, one can relate pairwise (rarely \mbox{one-to-many}) interactions of users with their declared social network. We describe the studies which tackle this issue in along the section.

\subsection{Activity networks versus declared social network}
The comparison of the network built from declared online friends and the network built from user interactions shows several differences at the structural level. First of all, the actors tend to interact with much less people than they declare as friends, what results in smaller degrees of nodes in the interactions network \cite{Viswanath2009Evolution, Wilson2009User}. Moreover, the friends they interact with change rapidly and only about $30\%$ of pairwise interactions in one month continue over the next month \cite{Viswanath2009Evolution}. Due to the fact that the degrees are lower, the properties related to small-world effect are also less evident, namely average path lengths are higher \cite{Wilson2009User} and there is less densely connected cores \cite{Chun2008Comparison}.

\subsection{Theories on social ties and information diffusion}

\begin{figure}\begin{center}
\includegraphics[width=12cm]{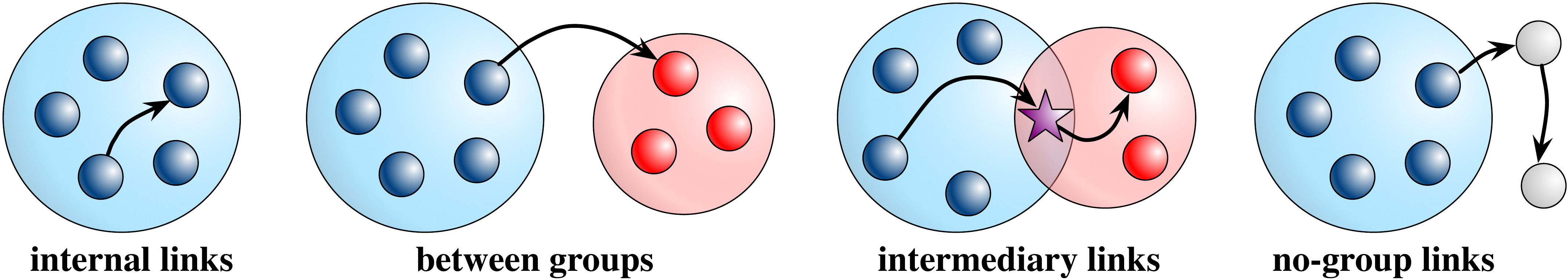}
\caption{ Different types of links depending on their position with respect to the groups' structure: internal, between groups, intermediary links and no-group links. Adapted from Ref.~\cite{Grabowicz2012Social}.}\label{fig_groups}
\end{center}\end{figure}

The theory known as {\it the strength of weak ties}, proposed by Granovetter~\cite{Granovetter1973strength}, deals with the relation between structure, intensity of social ties and diffusion of information in offline social networks. On one hand, a tie can be characterized by its strength, which is related to the time spend together, intimacy and emotional intensity of a relation. Strong ties refer to relations with close friends or relatives, while weak ties represent links with distant acquaintances. On the other hand, a tie can be characterized by its position in the network. Social networks are usually composed of communities. A tie can thus be internal to a group or a bridge between groups, as in Figure~\ref{fig_groups}. Granovetter's theory predicts that weak ties act as bridges between groups and are important for the diffusion of new information across the network. Strong ties are predicted to be located at the interior of the groups between actors who have many friends in common. Burt's work~\cite{Burt2005Brokerage} emphasizes the advantage of connecting different groups to access novel information due to the diversity in the sources.

Furthermore more recent works point out that information propagation may be dependent on the type of content transmitted~\cite{Centola2007Cascade, Centola2007Complex, Centola2010Spread} and on a \textit{diversity-bandwidth tradeoff}~\cite{Aral2011Diversity-Bandwidth}. The bandwidth of a tie is defined as the rate of information transmission per unit of time. Aral et al.~\cite{Aral2011Diversity-Bandwidth} note that weak ties interact infrequently, therefore have low bandwidth, whereas strong ties interact more often and have high bandwidth. The authors claim that both diversity and bandwidth are relevant for the diffusion of novel information. Since both are anticorrelated, there has to be a tradeoff to reach an optimal point in the propagation of new information. They also suggest that strong ties may be important to propagate information depending on the structural diversity, the number of topics and the dynamic of the information. Due to the different nature of online and offline interactions, it is not clear whether online networks organize following the sociological theories. In the following subsection we present results of some works testing if these theories apply to online social networks.

\subsection{Testing social theories in online social networks}
The predictions of the theory of {\it the strength of weak ties} have been checked in a mobile phone calls dataset~\cite{Onnela2007Structure} and, very recently, in online social networks~\cite{Grabowicz2012Social, Bakshy2012Role, Iribarren2011Affinity}.
Different predictors have been considered to estimate social tie strength~\cite{Marsden1984Measuring} including, for instance, time spent together~\cite{Marsden1984Measuring}, the duration of phone calls~\cite{Onnela2007Structure} or number of messages exchanged~\cite{Bakshy2012Role, Grabowicz2012Social}.
The two works~\cite{Onnela2007Structure, Grabowicz2012Social} have measured the dependence of strength of a tie on number of common friends shared by the two actors, showing that the more friends they share the more likely it is that the tie is strong. This stays in agreement with homophily effect in social network described at the beginning of this chapter. Many shared friends of a pair of users coupled by a strong tie can be interpreted as high homophily between them in terms of acquired friends. Furthermore, large field experiment performed at Facebook~\cite{Bakshy2012Role} has isolated the effect of homophily and social impact on the probability of propagation of information in online social network. The study has shown that users are around $7$-times more likely to re-broadcast a piece of information published by their friends if they are exposed to it, which is interpreted as $7$-times higher chance of information propagation due to social influence than to homophily. Moreover, the work argues that the weaker is the tie for which information propagation is considered, the higher is the likelihood of information flow due to social influence. This corresponds to Granovetter's prediction that weak ties are important for information diffusion. In the following paragraphs we describe in more detail findings of a similar study in Twitter \cite{Grabowicz2012Social}, a popular social microblogging platform.

\begin{figure}\begin{center}
\includegraphics[width=8cm]{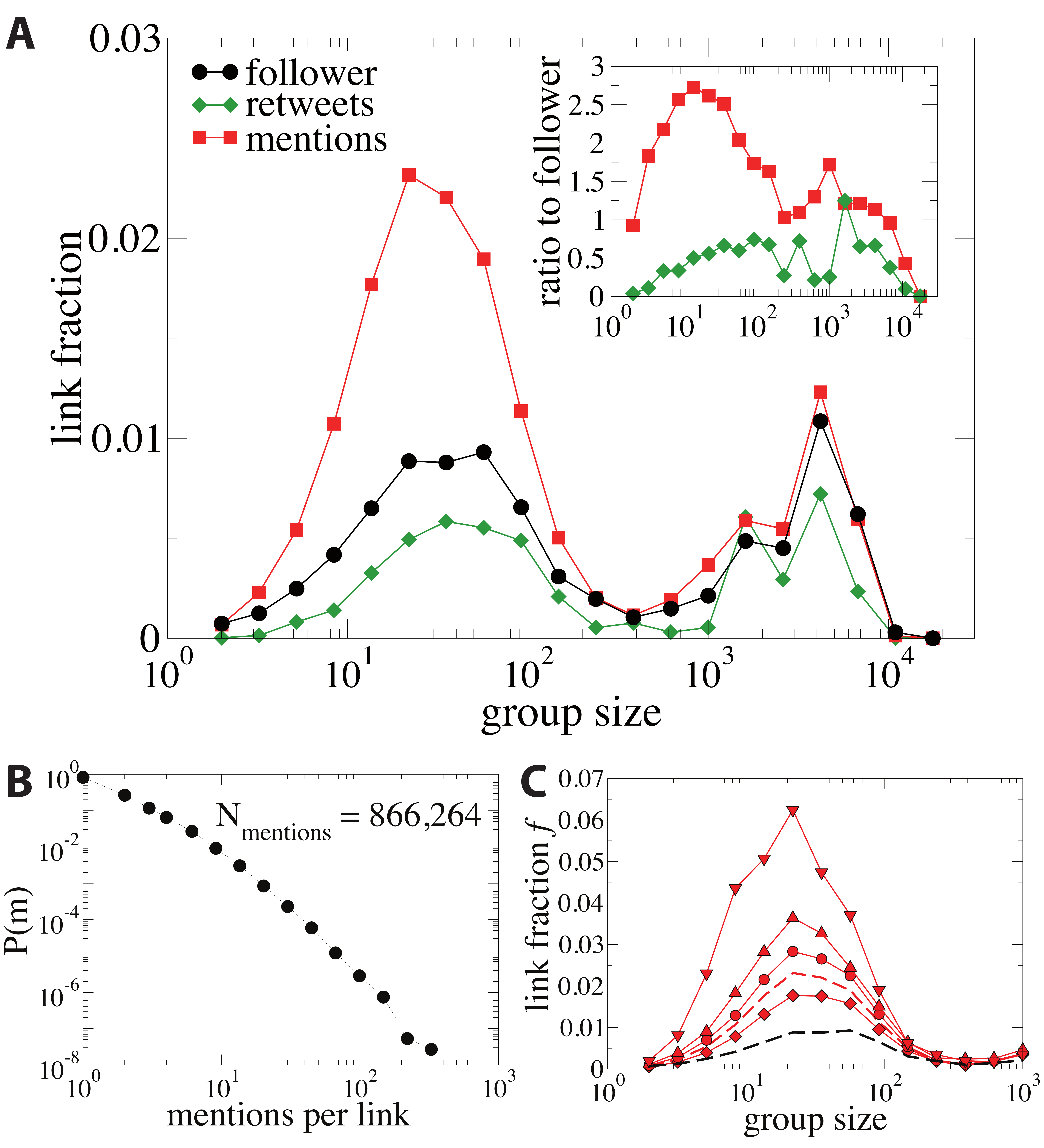}
\caption{Internal activity in Twitter. (A) Fraction $f$ of internal links as a function of the group size in number of users. The curve for the follower network acts as baseline (black) for mentions (red) and retweets (green). Note that if mentions/retweets were randomly appearing over follower links then the red/green curve should collapse with the black curve. Inset: links with mentions divided by the baseline (red), and links with retweets divided by the baseline (green). (B) Distribution of the number of mentions per link. (C) Fraction of links with mentions as a function of their intensity. The dashed curves are the total for the follower network (black) and for the links with mentions (red). While the other curves correspond (from bottom to top) to fractions of links with: 1 non-reciprocated mention (diamonds), 3 mentions (circles), 6 mentions (triangle up) and more than 6 reciprocated mentions (triangle down). From Ref.~\cite{Grabowicz2012Social}.} \label{fig_internal}
\end{center}\end{figure}

Online networks are promising for social studies due to the wide availability of data and the fact that different types of interactions are explicitly separated: e.g.,  information  diffusion events are distinguished from more personal communications. Diffusion events are implemented as a system option in the form of \textit{share}, or \textit{repost} buttons with which it is enough to single-click on a piece of information to rebroadcast it to all the users' contacts. This is in contrast to personal communications for which more effort has to be invested to write a short message and to select the recipient(s). In Twitter such actions are called respectively \textit{retweet}~\cite{Galuba2010Outtweeting} and \textit{mention/reply}~\cite{Honeycutt2009Beyond}.
The more mentions has been exchanged between two users, even more so if reciprocated, the stronger we consider the tie between them.
On the other hand declared network does also exist in Twitter and is made of directed follower links. One, using clustering algorithms, can find communities of more densely connected users in such network. Specifically, in the study which we present, various clustering algorithms have been used \cite{Grabowicz2012Supporting} and for brevity we will focus only on results for OSLOM~\cite{Lancichinetti2011Finding}. Granovetter theory predicts that social ties should occur more often inside communities. This is what happens for links with mentions.
We define the fraction $f$ as ratio between the number of links with specified type of interaction in given position with respect to the groups of corresponding size and the total number of links with that interaction. The fractions $f$ reveals an interesting pattern as function of the group size as can be seen in Figure~\ref{fig_internal}A. Links with mentions are more abundant inside communities than any other links. This effect is especially significant for groups of sizes from $10$ to $150$ members. In addition, the distribution of the number of times that a link is used (intensity) for mentions is wide, which allows for a systematic study of the dependence of intensity and position (see Figure~\ref{fig_internal}B). It turns out that the more intense (or reciprocated) a link with mentions is, the more likely it becomes to find this link as internal (Figure~\ref{fig_internal}C). This corresponds to Granovetter expectation that the stronger the tie is the higher the number of mutual contacts of both parties it has and the higher the chance that the parties belong to the same group.

\begin{figure}\begin{center}
\includegraphics[width=12cm]{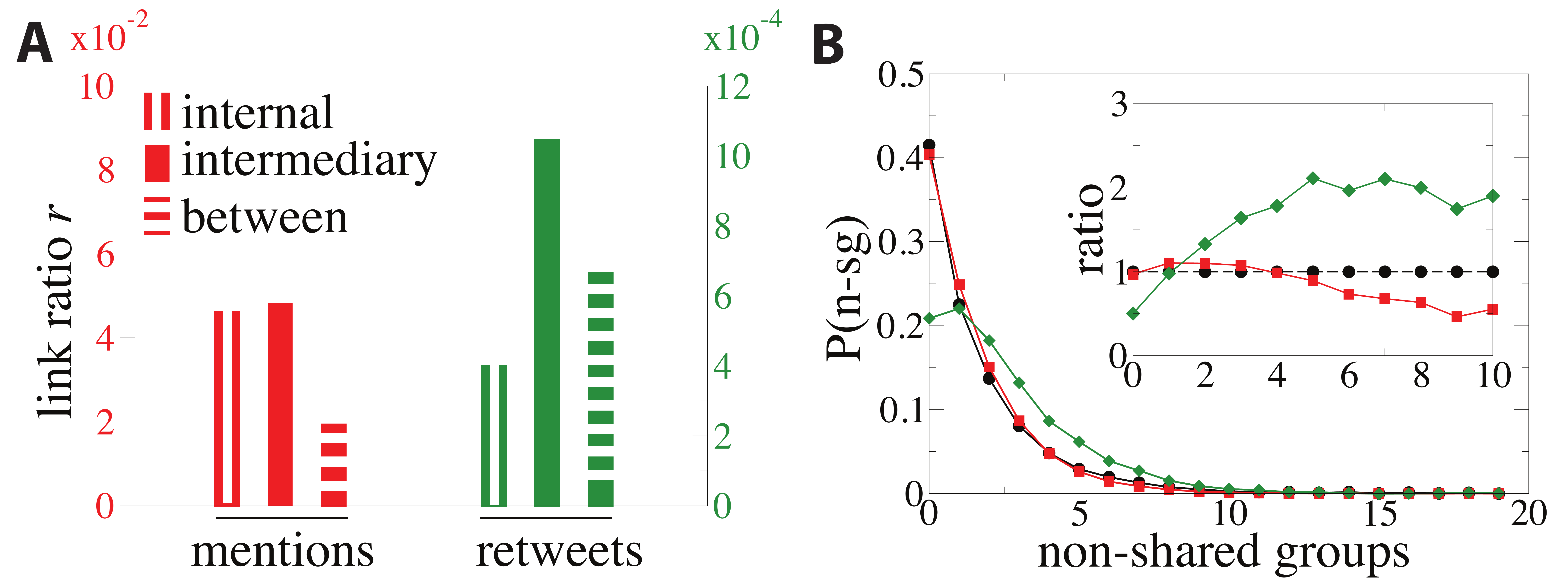}
\caption{Intermediary links. (A) Ratio $r$  between the number of links with mentions or retweets and number of follower links. (B) Distribution of the links in the follower network (black curve), those with mentions (red curve) and retweets (green curve) as a function of the number of non-shared groups of the users connected by the link. Inset, ratios between these distributions and the follower network. From Ref.~\cite{Grabowicz2012Social}.} \label{fig_intermediary}
\end{center}\end{figure}

The communication between groups can take place in two ways: the information can propagate by means of links between groups or by passing through an intermediary user belonging to more than one group, see Figure~\ref{fig_groups}. We have defined as intermediary the links connecting a pair of users sharing a common group and with at least one of the users belonging also to a different group.
In order to estimate the efficiency of the different types of links as attractors of mentions and retweets, there was measured a ratio $r$ defined as number of links with specified interaction in a given position divided by the total number of links in that position.
The bar plot with the values of $r$ is displayed in Figure~\ref{fig_intermediary}. The efficiency of the different type of links can thus be compared for the attraction of mentions (red bars) and retweets (green bars). Links internal to the groups attract more mentions and less retweets than links between groups in agreement with the predictions of the strength of weak ties theory. Intermediary links attract mentions as likely as internal links: the ratio of intermediary links with mentions is very close to the ratio of internal links with mentions. This is expected because intermediary links are also internal to the groups. However, the aspect that differentiates more intermediary links from other type of links is the way that they attract retweets. Intermediary links bear retweets with a higher likelihood than either internal or between-groups connections (see Figure~\ref{fig_intermediary}A). This fact can be interpreted within the framework of the tradeoff between diversity and bandwidth~\cite{Aral2011Diversity-Bandwidth}: strong ties are expected to be internal to the groups and to have high bandwidth, while ties connecting diverse environments or groups are more likely to propagate new information. High bandwidth links in our case correspond to those with multiple mentions, while links providing large diversity are the ones between groups. Intermediary links exhibit these two features: they are internal to the groups and statistically bear more mentions, and introduce diversity through the intermediary user membership in several groups. Moreover, in line with the theories~\cite{Granovetter1973strength, Burt2005Brokerage, Aral2011Diversity-Bandwidth}, higher diversity increases the chances for a link to bear retweets as can be seen in Figure~\ref{fig_intermediary}B, which implies a more efficient information flow. In the inset of the Figure it is shown that the number of non-shared groups assigned to the users connected by the link positively correlates with, up to twice higher than expected, number of retweets.

\section{Summary}
Research of online social networks is a rich and an active field of study. The availability of large amount of data allows for studies of both dynamics of social networks and user-user activity on the social network connections. Different growth models have been proposed to simulate the growth of the network, among which three main families are: preferential growth models, fitness or hidden variables  models and triangle-closing models. The latter model is reported to yield most accurate results; however, it also incorporates mechanism of preferential attachment. The main advantage of triangle-closing model is that it directly produces network with enough clustering, which is reported to be a feature of social networks. Moreover, there are still open questions about the origin of these mechanisms and of some other phenomena observed during the growth process such as network densification \cite{Leskovec2007Graph}. While declared social network evolves different types of interactions occur between its members, mostly between users already connected in the declared social network.

Recent studies have shown that different types of interactions happen according to the patterns predicted by the sociological theories. In general strong ties, which in online social networks are usually defined as the links with many messages exchanged between the pair of users, happen more often between users who have many friends in common, or who belong to the same communities. On the other hand, weak ties appear more often between users who do not share friends and belong to different groups. It has been shown that weak ties are more efficient for the information spreading than strong ties. Closer study shows that tradeoff between diversity and bandwidth may be crucial for diffusion of information.

In conclusion dynamics and activity in online social networks is remarkably rich and tells us much about our social behavior and confirms some of the known offline social theories. We expect that this field of research will be active and developing in the following years and that numerous further online observations and experiments will be undertaken to better understand and quantitatively describe social behaviors.

\section{Acknowledgments}
We acknowledge partial support from the European Commission through PATRES project, the Spanish Ministry of Economy (MINECO) and FEDER (EU) through projects FISICOS (FIS2007-60327) and MODASS (FIS2011-247852). P.A.G. acknowledges support from the JAE program of the CSIC; J.J.R. acknowledges support from the Ram\'on y Cajal program of MINECO.

%
%



\printindex
\end{document}